\documentstyle[12pt]{article}
\begin{document}
\thispagestyle{empty}
\begin{center}
\renewcommand{\baselinestretch}{1.5}{\LARGE \tt \bf Riemannian geometrical constraints on magnetic vortex filaments in plasmas}
\end{center}
\vspace{0.5cm}
\begin{center}
{\large \tt \bf L. C. Garcia de Andrade \footnote{Departamento de F\'{\i}sica Teorica, Instituto de F\'{\i}sica,UERJ,Brasil-garcia@dft.if.uerj.br}}
\end{center}
\vspace{2cm}
\begin{abstract}
  Two theorems on the Riemannian geometrical constraints on vortex magnetic filaments acting as dynamos in (MHD) flows are presented. The use of Gauss-Mainard-Codazzi equations allows us to investigate in detail the influence of curvature and torsion of vortex filaments in the MHD dynamos. This application follows closely previous applications to Heisenberg  spin equation to the investigations in magnetohydrostatics given by Schief (Plasma Physics J. 10, 7, 2677 (2003)). The Lorentz force on vortex filaments are computed and the ratio between the forces along different directions are obtained in terms of the ratio between the corresponding magnetic fields which equals also the ratio between the Frenet torsion and vortex line curvature. A similar relation between Lorentz forces, magnetic fields and twist, which is proportional to total torsion integral has been  obtained by Ricca (Fluid Dyn. Res. 36,319 (2005)) in the case of inflexional desiquilibrium of magnetic flux-tubes. This is due to the fact that the magnetic vortex lines are a limit case of the magnetic flux-tubes when the lenght of the tube is much greater than the radius of the tube. Magnetic helicity equation of the filament allows us again to determine the magnetic fields ratio from Frenet curvature and torsion of the vortex lines. 
\end{abstract}
\noindent
\hspace{2cm} 
\section{Introduction}
 Recently Schief  \cite{1} have shown  that the classical magnetohydrostatic equations of infinitely conducting fluids may be reduced to the integral potential Heisenberg equation constraint to a Jacobian condition as long as the magnetic field is constant along individual magnetic lines. Palumbo's \cite{2} toroidal isodynamic equilibrium has been given as an  example. Earlier Schief has also \cite{3} had provided another very interesting application of how the use of curvature and torsion of lines affects the plasma physical phenomena, by showing that the equilibrium equations of MHD reduce to integral Pohlmeyer-Lund-Regge \cite{4} model subject to a volume preserving constraint if Maxwellian surfaces are assumed to coincide with total pressure constant surfaces. In that paper he provided nested toroidal flux surfaces in magnetohydrostatics. In this paper we provide a new application of the use of the mathematical machinery developed by Rogers and Schief \cite{6} to plasma physics. Namely , we apply the Gauss-Mainardi-Coddazzi equations (GMC) and the Heisenberg spin  \cite{7} to MHD dynamos \cite{8} to compute curvature and torsion effects on vortex filaments of the magnetic field lines. Torsion effects on vortex filaments with and without magnetic fields have been previously investigated by Ricca \cite{9}. More recently Garcia de Andrade \cite{10} have investigated the equilibrium of the magnetic stars more well-known as magnetars. Another interesting example to plasma physics is provided by the Beltrami magnetic flows \cite{11}. These are very important problems in plasma physics and therefore new mathematical methods to address the problem may shed light on their solutions and their applications. This investigation seems to be useful to mathematical and plasma physicists. The paper is organised as follows: Section II we review the mathematics apparatus of the Serret-Frenet equations and the Heisenberg spin equations. In section III we investigate the application of this mathematical framework in explicitely plasma physics problems as the Beltrami magnetic dynamos and the effects of curvature and torsion on vortex filaments. Section IV we compute the Lorentz force on the vortex filaments and in section V we discuss the Beltrami fields while finally in section VI we present discussions and conclusions.  
\section{Geometrical constraints on the magnetic vortex filaments}
 In this section we reproduce for the benefit of the reader some of the formulas derived by Rogers and Schief \cite{6} on the Heisenberg spin equation and geometry of curvature and torsion of lines. We begin by defining a Serret-Frenet frame composed of the vectors triad  $X=(\vec{t},\vec{n},\vec{b})$. The extended Serret-Frenet formulas can be written in matrix form are given by
\begin{equation}
\frac{\partial}{{\partial}s}X^{T}= A X^{T}
\label{1}
\end{equation}
where A is given by the array
\vspace{1.0cm}
\[\displaylines{\pmatrix{0&\kappa&0\cr
	-\kappa&0&\tau\cr
	0&-\tau &0\cr}\cr}\]

while the other equations for $\vec{n}$ and $\vec{b}$ direction are given
\begin{equation}
\frac{\partial}{{\partial}n}X^{T}= B X^{T}
\label{2}
\end{equation}
\begin{equation}
\frac{\partial}{{\partial}b}X^{T}= C X^{T}
\label{3}
\end{equation}
where T here represents the transpose of the line matriz X and B and C are the respective skew-symmetric matrices
\vspace{1.0cm}
\[\displaylines{\pmatrix{0&{\theta}_{ns}&{\Omega}_{b}+{\tau}\cr
	-{\theta}_{ns}&0&-div\vec{b}\cr
	-({\Omega}_{b}+{\tau})&div\vec{b} &0\cr}\cr}\]
and
\vspace{1.0cm}\[
\displaylines{\pmatrix{0&-({\Omega}_{n}+{\tau})&{\theta}_{bs}\cr
	({\Omega}_{n}+{\tau})&0&\kappa+div\vec{n}\cr
	-{\theta}_{bs}&-(\kappa+div\vec{n}) &0\cr}\cr}\]

where ${\theta}_{ns}$ and ${\theta}_{bs}$ are respectively given by
\begin{equation}
{\theta}_{ns}=\vec{n}.\frac{\partial}{{\partial}n}\vec{t}
\label{4}
\end{equation}
and
\begin{equation}
{\theta}_{bs}=\vec{b}.\frac{\partial}{{\partial}b}\vec{t}
\label{5}
\end{equation}
The gradient operator is 
\begin{equation}
{\nabla}= \vec{t}\frac{\partial}{{\partial}s}+\vec{n}\frac{\partial}{{\partial}n}+\vec{b}\frac{\partial}{{\partial}b}
\label{6}
\end{equation}
The other vector analysis formulas read
\begin{equation}
div\vec{t}={\theta}_{ns}+{\theta}_{bs}
\label{7}
\end{equation}
\begin{equation}
div\vec{n}= -{\kappa}+\vec{b}.\frac{\partial}{{\partial}b}\vec{n}
\label{8}
\end{equation}
\begin{equation}
div\vec{b}= -\vec{b}.\frac{\partial}{{\partial}n}\vec{n}
\label{9}
\end{equation}
\begin{equation}
{\nabla}{\times}\vec{t}={\Omega}_{s}\vec{t}+ \kappa\vec{b}
\label{10}
\end{equation}
where
\begin{equation}
{\Omega}_{s}=\vec{b}.\frac{\partial}{{\partial}n}\vec{t}-\vec{n}.\frac{\partial}{{\partial}b}\vec{t}
\label{11}
\end{equation}
which is called abnormality of the ${\vec{t}}$-field. Similarly the results for ${\vec{n}}$ and ${\vec{b}}$ are given  by
\begin{equation}
{\nabla}{\times}\vec{n}= -(div{\vec{b}})\vec{t}+ {\Omega}_{n}\vec{n}
\label{12}
\end{equation}
\begin{equation}
{\Omega}_{n}=\vec{n}.{\nabla}{\times}\vec{n}=-\vec{t}.\frac{\partial}{{\partial}b}\vec{n}-\tau
\label{13}
\end{equation}
and
\begin{equation}
{\nabla}{\times}\vec{b}= ({\kappa}+div{\vec{n}})\vec{t}-{\theta}_{bs}+{\Omega}_{b}\vec{b}
\label{14}
\end{equation}
\begin{equation}
{\Omega}_{b}=\vec{b}.{\nabla}{\times}\vec{b}=-\vec{t}.\frac{\partial}{{\partial}n}\vec{b}-\tau
\label{15}
\end{equation}
where
\begin{equation}
{\Omega}_{s}=\vec{b}.\frac{\partial}{{\partial}n}\vec{t}-\vec{n}.\frac{\partial}{{\partial}b}\vec{t}
\label{16}
\end{equation}
which is called abnormality of the ${\vec{t}}-field$. Similarly the results for ${\vec{n}}$ and ${\vec{b}}$ are given  by
\begin{equation}
{\nabla}{\times}\vec{n}= -(div{\vec{b}})\vec{t}+ {\Omega}_{n}\vec{n}
\label{17}
\end{equation}
\begin{equation}
{\Omega}_{n}=\vec{n}.{\nabla}{\times}\vec{n}=-\vec{t}.\frac{\partial}{{\partial}b}\vec{n}-\tau
\label{18}
\end{equation}
and
\begin{equation}
{\nabla}{\times}\vec{b}= ({\kappa}+div{\vec{n}})\vec{t}-{\theta}_{bs}+{\Omega}_{b}\vec{b}
\label{19}
\end{equation}
\begin{equation}
{\Omega}_{b}=\vec{b}.{\nabla}{\times}\vec{b}=-\vec{t}.\frac{\partial}{{\partial}n}\vec{b}-\tau
\label{20}
\end{equation}
To simplify the magnetic field computations in the next section we shall consider here the particular case of ${\Omega}_{n}=0$ which as has been shown by Rogers and Schief implies the complex lamelar motions and the constancy of magnitude along the streamlines. This geometrical condition implies that the existence of two scalar functions ${\Phi}$ and ${\psi}$ which satisfy the relation
\begin{equation}
\vec{n}={\psi}{\nabla}{\Phi}
\label{21}
\end{equation}
Since tangent planes to the surfaces ${\Phi}=constant$ are generated by the unit tangent $\vec{t}$ and the binormal $\vec{b}$, or  
\begin{equation}
\vec{t}.{\nabla}\Phi= 0
\label{22}
\end{equation}
and
\begin{equation}
\vec{b}.{\nabla}\Phi= 0
\label{23}
\end{equation}
Since $\vec{n}$ is parallel to the normal to surfaces ${\Phi}=const$, the vector lines $\vec{t}$ are geodesics on the surfaces which implies taht the $b-lines$ are parallels on the surface ${\Phi}=const$. The s-lines and b-lines being the parametric curves on the ${\Phi}=const$ surface then a surface metric can be written as 
\begin{equation}
I=d{s^{2}}+g(s,b)db^{2}
\label{24}
\end{equation}
In accordance with the Gauss-Weingarten equations for ${\Phi}=const$ we have the same Serret-Frenet matriz above and  
\begin{equation}
\frac{1}{g^{\frac{1}{2}}}\frac{\partial}{{\partial}b}X^{T}= D X^{T}
\label{25}
\end{equation}
where the matrix D is
\vspace{1.0cm}
\[\displaylines{\pmatrix{0&-{\tau}&{\theta}_{bs}\cr
	{\tau}&0&\kappa+div\vec{n}\cr
	-{\theta}_{bs}&-(\kappa+div\vec{n}) &0\cr}\cr}\]
As shown by Rogers and Schief the $\vec{t}-field$ satisfies the Heisenberg spin-type equation
\begin{equation}
\frac{\partial}{{\partial}b}\vec{t}=\frac{\partial}{{\partial}s}(h\vec{t}{\times}\frac{\partial}{{\partial}s}\vec{t})
\label{26}
\end{equation}
where $h=\frac{g^{\frac{1}{2}}}{\kappa}$. The Gauss-Mainardi-Codazzi equations are 
\begin{equation}
{g^{\frac{1}{2}}}\frac{\partial}{{\partial}b}\kappa+\frac{\partial}{{\partial}s}(g\tau)=0
\label{27}
\end{equation}
\begin{equation}
\frac{\partial}{{\partial}b}{\tau}=\frac{\partial}{{\partial}s}[{g^{\frac{1}{2}}}(\kappa+div\vec{n})]+\kappa\frac{\partial}{{\partial}s}{g^{\frac{1}{2}}}
\label{28}
\end{equation}
\begin{equation}
{g^{\frac{1}{2}}}[{\kappa}({\kappa}+div\vec{n})+{\tau}^{2}]=\frac{{\partial}^{2}}{{\partial}s^{2}}{g^{\frac{1}{2}}}
\label{29}
\end{equation}
Besides Rogers and Schief also showed that the Heisenberg spin equation implies the relation
\begin{equation}
\frac{\partial}{{\partial}s}{\kappa}= {\kappa}{\theta}_{bs}
\label{30}
\end{equation}
Most of the expressions revised in this section would be used on the next section in the derivation  of the magnetic field dynamo equations in the Serret-Frenet frame.
\section{Geometrical constraints on MHD dynamos and magnetic helicity}
We start tn this section with a very simple theorem concerning geometrical constraints in the Salingaros \cite{11} formula for the self-exciting MHD dynamos phenomenologically based, which is expressed as
\begin{equation}
{\nabla}{\times}\vec{B}=k\vec{v}{\times}\vec{B}
\label{31}
\end{equation}
where the magnetic field in general is $\vec{B}=B(s,b,n)\vec{t}$.
\newpage 
THEOREM 1: A vortex magnetic filament maybe act physically as a Salingaros MHD dynamo where the magnetic field and vortex filament velocity are given by $\vec{B}=B(s)\vec{t}$ and $\vec{v}={\kappa}\vec{b}$ respectively as long as the vortex filament is represented by a straight (curvature $\kappa=0$) vortex line and the hydrodynamical flow undergoes geodesic motions. Besides the modulus of the magnetic field is constant along the straight vortex filament.
Proof: Let us consider the expansion of the LHS of equation (\ref{31}) as  
\begin{equation}
B{\nabla}{\times}\vec{t}+{\nabla}B{\times}\vec{t}=B({\Omega}_{s}\vec{t}+\kappa\vec{b})=-kB\vec{n}
\label{32}
\end{equation}
Note that this equation yields the simple equations
\begin{equation}
B\kappa=0
\label{33}
\end{equation}
\begin{equation}
B({\Omega}_{s}+\kappa h{\tau})=0
\label{34}
\end{equation}
Since the modulus of the magnetic field B is by assumption nonvanishing, equations (\ref{33}) and (\ref{34}) together yields $\kappa=0$ and ${\Omega}_{s}=0$. The first contraints the vortex filament to straight line while the second tells us that the hydrodynamical motion is geodesic. The last part of the theorem $1$ is easily proved from the other Maxwell equation 
\begin{equation}
{\nabla}.\vec{B}= 0
\label{35}
\end{equation}
which imply 
\begin{equation}
\frac{\partial}{{\partial}s}{B}= 0
\label{36}
\end{equation}
A similar theorem for the magnetic helicity shall be considered now where a more general type of magnetic field and vortex filament velocity shall be considered. Of course this can also beconsidered in the case of phenomenological dynamos.
\newpage
THEOREM 2: The magnetic helicity equation ${\nabla}{\times}\vec{B}={\lambda}\vec{B}$ allows us to write the ratio between the magnetic fields along the s and b-lines in terms of geometrical quantities such as Frenet curvature and torsion of the vortex filament as long as we consider the magnetic field of the filament in the form $\vec{B}=B_{s}\vec{t}+ B_{b}\vec{b}$. The hydrodynamical flows undergo geodesic motions also in this case. Proof:
Let us consider the expansion of the LHS of the magnetic helicity \cite{12}
\begin{equation}
B_{s}{\nabla}{\times}\vec{t}+{\nabla}B_{s}{\times}\vec{t}+B_{b}{\nabla}{\times}\vec{t}+{\nabla}B_{b}{\times}\vec{t}=k(B_{s}\vec{t}+B_{b}\vec{b})
\label{37}
\end{equation}
A long but straightforward computation leads to the following set of the equations
\begin{equation}
\frac{\partial}{{\partial}s}B_{b}-\frac{\partial}{{\partial}b}B_{s}-{\theta}_{bs}B_{b}[1-hk]=0
\label{38}
\end{equation}
\begin{equation}
2\frac{B_{s}}{B_{b}}\kappa+[{\Omega}_{b}+{\tau}]={\lambda}
\label{39}
\end{equation}
\begin{equation}
(\kappa+div\vec{n})\frac{B_{b}}{B_{s}}+[h\tau\kappa+{\Omega}_{s}]={\lambda}
\label{40}
\end{equation}
To end the proof let us consider some particular cases as for example the straight vortex lines consider in the first theorem. In this case $\kappa=0$ and $\tau=0$ therefore the system of equations reduce to
\begin{equation}
{\Omega}_{b}={\lambda}
\label{41}
\end{equation}
\begin{equation}
(div\vec{n})\frac{B_{b}}{B_{s}}+{\Omega}_{s}={\lambda}
\label{42}
\end{equation}
These two equations together imply that the ratio of the magnetic fields are given by
\begin{equation}
\frac{B_{b}}{B_{s}}=-\frac{[{\Omega}_{s}-{\Omega}_{b}]}{div\vec{n}}
\label{43}
\end{equation}
when the geodesic motions of flow are allowed one has ${\Omega}_{s}=0$ and this expression can be further simplified. When the vortex filament is not straight one may consider that the torsion vanishes and the filament is a planar curve where $\kappa=constant$. By the Gauss-Weingarten equations and the Heisenberg spin equations the expression (\ref{30}) implies ${\theta}_{bs}=0$. 
In this equation (\ref{38}) reduces to 
\begin{equation}
\frac{\partial}{{\partial}s}B_{b}-\frac{\partial}{{\partial}b}B_{s}=0
\label{44}
\end{equation}
along with the Maxwell equation
\begin{equation}
{\nabla}.\vec{B}=\frac{\partial}{{\partial}s}B_{s}+\frac{\partial}{{\partial}b}B_{b}=0
\label{45}
\end{equation}
Together equations (\ref{44}) and (\ref{45}) yields
\begin{equation}
B_{b}=\frac{\partial}{{\partial}b}{\Phi}
\label{46}
\end{equation}
\begin{equation}
B_{s}=\frac{\partial}{{\partial}s}{\Phi}
\label{47}
\end{equation}
and 
\begin{equation}
\frac{{\partial}^{2}}{{\partial}s^{2}}{\Phi}+\frac{{\partial}^{2}}{{\partial}b^{2}}{\Phi}=0
\label{48}
\end{equation}
which is a Laplacian like formal equation. Note also that this time the field $\vec{B}(s,b)$ has no constant modulus along the vortex line. In general one may notice that for general nonuniform torsion and curvature the ration between the magnetic fields becomes
\begin{equation}
\frac{B_{b}}{B_{s}}=-\frac{2\kappa}{[({\Omega}_{b}+\tau)-{\lambda}]}
\label{49}
\end{equation}
which shows clearly that the  ratio of the magnetic fields of the vortex filaments along different directions can be expressed in terms of the geometrical quantities $\kappa$ and $\tau$.
Actually in the next section when we compute the ratio between the Lorentz force components along the lines s and b, we shall see that similar constraints appear between this ratio and the ratio obtained here of the magnetic fields. The only problem is that for magnetic helicity field the Lorentz force vanishes since $\vec{F}=[{\nabla}{\times}\vec{B}]{\times}\vec{B}={\lambda}\vec{B}{\times}\vec{B}=0$.
\section{Lorentz force on vortex filaments in MHD} 
Very recently Ricca \cite{8} has shown that magnetic flux-tubes in inflexional configuration are in equilibrium evolving to inflexion-free state. He showed that by deriving the Lorentz force associated with the magnetic flux-tube one could use them for investigate the generic behaviour associated with the passage to the inflexional configuration. His investigation has proved to be useful in applications to solar corona loops and astrophysical flows. In this section by reasoning that the magnetic vortex filaments are limiting case of magnetic flux tubes we show that similar geometrical constraints are imposed to the magnetic vortices. Let us now consider the Lorentz equation
\begin{equation}
\vec{F}=[{\nabla}{\times}\vec{B}]{\times}\vec{B}= k\vec{v}{\times}\vec{B}	
\label{50}
\end{equation}
Taking the velocity vector as $\vec{v}=v_{s}\vec{t}+v_{b}\vec{b}$ into the expression  (\ref{50}) one obtains
\begin{equation}
F_{s}\vec{t}+F_{b}\vec{b}= -k[(v_{s}B_{b}-v_{b}B_{s})\vec{b}+(v_{b}B_{s}-v_{s}B_{b})\vec{t}]
\label{51}
\end{equation}
which yields  the following ratio between the Lorentz force components along different directions
\begin{equation}
\frac{F_{b}}{F_{s}}= \frac{B_{s}}{B_{b}}
\label{52}
\end{equation}
Expansion on the LHS of equation (\ref{50}) allows us to obtain the following equations 
\begin{equation}
\frac{B_{s}}{B_{b}}= [\frac{v_{s}}{v_{b}}-\frac{{\theta}_{bs}}{{\kappa}{v}_{b}}
\label{53}
\end{equation}
Due to the expression
\begin{equation}
{\theta}_{bs}=\frac{\partial}{{\partial}s}\kappa
\label{54}
\end{equation}
Besides the magnetic field components can be expressed in terms of Frenet torsion and curvature of the vortex filaments as
\begin{equation}
B_{s}= \frac{\tau}{\kappa}
\label{55}
\end{equation}
\begin{equation}
B_{b}= \frac{\tau}{\kappa}[\frac{v_{s}}{v_{b}}-{\theta}_{bs}]
\label{56}
\end{equation}
Note from this expression that when the curvature is constant as happens along the helical vortices or flows the ${\theta}_{bs}$ vanishes and $B_{b}$ reduces to
\begin{equation}
B_{b}= \frac{\tau}{\kappa}[\frac{v_{s}}{v_{b}}]
\label{57}
\end{equation}
Now is easy from these expressions to obtain the relation  between the  Lorentz force components 
\begin{equation}
\frac{F_{b}}{F_{s}}= \frac{B_{s}}{B_{b}}= [\frac{v_{s}}{v_{b}}-\frac{1}{v_{b}{\kappa}}\frac{\partial}{{\partial}s}\kappa]
\label{58}
\end{equation}
Note that when the motion of the vortex filament is only along the b-direction, which is the most usual situation \cite{13}  $v_{s}$ vanishes and (\ref{58}) reads  
\begin{equation}
\frac{F_{b}}{F_{s}}= -{\frac{\partial}{{\partial}s}\kappa}{{\kappa}^{2}}
\label{59}
\end{equation}
This expression can be recast in a more interesting form if we notice that the Gauss-Mainardi-Codazzi equation (\ref{27}) for $\kappa=\kappa(s)$ can be expressed as
\begin{equation}
\frac{\partial}{{\partial}s}(g\tau)=0
\label{60}
\end{equation}
which yields $g\tau=c(b)$ where c is an arbitrary constant for the s variable. Since $g={\kappa}^{2}h^{2}$ one may express the torsion  $\tau$ in terms of curvature $\kappa$ as
\begin{equation}
\frac{c}{\tau{h^{2}}}={\kappa}^{2}
\label{61}
\end{equation}
Substitution of ${\kappa}^{2}$ into (\ref{61}) yields the final expression
\begin{equation}
\frac{F_{b}}{F_{s}}= \frac{{\tau}h^{2}{\theta}_{bs}}{R(s)}
\label{62}
\end{equation}
where $R(s)$ is the radius of curvature where $\kappa(s)=\frac{1}{R}$. Just for comparison we repeat Ricca's expression here
\begin{equation}
\frac{F_{s}}{F_{\theta}}= -\frac{2{\pi}{T{\omega}r}}{LK(s)}
\label{63}
\end{equation}
where $T{\omega}$ is the twist which is proportional to the total torsion $\frac{1}{2\pi}\int{{\tau}ds}$. This makes our expression similar to Ricca's one for the magnetic flux-tube.
\section{Beltrami fields and flows}
Beltrami magnetic flows are given by 
\begin{equation}
{\nabla}{\times}\vec{v}=k\vec{v}
\label{64}
\end{equation}
Let us consider that the flow velocity is along the s-line direction or $\vec{v}=v_{t}\vec{t}$. One may note that the LHS of equation (\ref{64}) represents the vorticity $\vec{\omega}$ which from Rogers and Schief work \cite{10} yields 
\begin{equation}
\vec{\omega}=v_{t}{\Omega}_{s}\vec{t}+(\frac{\partial}{{\partial}b}v_{t})\vec{n}+(v_{t}{\kappa}-\frac{\partial}{{\partial}n}v_{t})\vec{b}
\label{65}
\end{equation}
Substitution of (\ref{65}) into the Beltrami flow equation yields
\begin{equation}
v_{t}{\Omega}_{s}\vec{t}+\frac{\partial}{{\partial}b}v_{t}\vec{n}+(v_{t}{\kappa}-\frac{\partial}{{\partial}n}v_{t})\vec{b}=m(v_{t}\vec{t})
\label{66}
\end{equation}
Note that the this vectorial equation yields three equations
\begin{equation}
{\Omega}_{s}= m=constant
\label{67}
\end{equation}
\begin{equation}
\frac{\partial}{{\partial}b}v_{t}=0
\label{68}
\end{equation}
\begin{equation}
\frac{\partial}{{\partial}n}v_{t}= {\kappa}v_{t}
\label{69}
\end{equation}
Note that the equation (\ref{67}) is differnt from the equation in the case of MHD dynamo phenomenology proposed by Salingaros, since the abnormality ${\Omega}_{s}$ does not vanish othewise the Beltrami flow would be irrotational or ${\omega}=0$. Applying the divergence operator to the Beltrami flow equation yields
\begin{equation}
{\nabla}.{\nabla}{\times}\vec{v}={\nabla}.\vec{v}
\label{70}
\end{equation}
Since the LHS of the equation (\ref{53}) vanishes the RHS yields
\begin{equation}
{\nabla}.\vec{v}=0
\label{71}
\end{equation}
implies
\begin{equation}
\frac{\partial}{{\partial}s}v_{s}=0
\label{72}
\end{equation}
which shows that $v_{t}$ only depends on the n-line direction. Let us now compute the Beltrami magnetic field vortex line. By analogous computations we did in the case of Beltrami flows yields 
\begin{equation}
\frac{\partial}{{\partial}b}B=0
\label{73}
\end{equation}
and
\begin{equation}
\frac{\partial}{{\partial}n}B+{\kappa}B=0
\label{74}
\end{equation}
along with the equation of no magnetic monopole
\begin{equation}
{\nabla}.\vec{B}=0
\label{75}
\end{equation}
yields
\begin{equation}
\frac{\partial}{{\partial}s}B=0
\label{76}
\end{equation}
which shows that Beltrami field only depends on the n-direction. This is amore general solution than the one obtained by Salingaros.
\section{Conclusions} 
 In conclusion, the effects of curvature and torsion are displayed in MHD dynamos by making use of solitonic Heisenberg spin-like equation and the Gauss-Mainardi-Codazzi equations. Magnetic helicity fields are also investigated and we show that the curvature and torsion effects are similar to the ones obtained  by Ricca for the magnetic flux-tubes. The comprehension of the geometry and dynamics of vortex filaments in MHD dynamos maybe certainly useful in applications to astrophysical plasma and solar physics \cite{14}. 
\section*{Acknowledgements}
I would like to thank CNPq (Brazil) for financial support as well a
Universidade do Estado do Rio de Janeiro for financial support. 


\newpage
\begin{thebibliography}{14}
\bibitem{1} W.K. Schief, J. Plasma Physics, 65,6,465 (2003). 
\bibitem{2} D. Palumbo, Nuovo Cimento B 53,507 (1968). 
\bibitem{3} W.K. Schief, Physics of Plasmas 10,7, 2677 (2003). 
\bibitem{4} K. Pohlmeyer, Comm. Math. Phys.46 ,207 (1976). F. Lund and T. Regge, Phys. Review D14, 1524 (1976). 
\bibitem{5} C. E. Weatherburn, Differential Geometry of Three Dimensions, Vol. II,Cambridge University Press, Cambridge UK, (1930). 
\bibitem{6} C. Rogers and Schief,J. Math. Analysis and Appl. 251, 855 (2000).
\bibitem{7} R. J. Goldston and P. H. Rutherford, Introduction to Plasma Physics, Institute of Physics, (2001) UK.
\bibitem{8} R. Ricca, Inflexional disequilibrium of magnetic flux-tubes ,Fluid Dynamics Research 36 (2005) 319. R. Ricca, Evolution and Inflexional stability of twisted magnetic flux tubes. Solar Physics 172 (1997) 241. R. Ricca, Phys. Rev. A (1999). 
\bibitem{9} R. Ricca, The effect of torsion on the motion of a helical vortex filament, Journal of Fluid Mechanics, 237, 241 (1994).
\bibitem{10} L.C. Garcia de Andrade, Curvature and Torsion effects on vortex filaments and Hasimoto soliton transformation, astro-ph/0509847 (2005).
\bibitem{11} S. Salingaros, Phys. Lett. A 185, 201 (1994).
\bibitem{12} M. A. Berger, Phys. Rev. Lett.70,6,705 (1993).
\bibitem{13} P.G. Saffaman, Vortex Dynamics,Cambridge University Press (2001).
\bibitem{14} J. Brat et al, Plasma Loops in the Solar Corona , Cambridge University Press (1991).
\end{thebibliography}
\end{document}